\documentclass[%
superscriptaddress,
reprint,
showpacs,
 amsmath,amssymb,
 aps,
 prl
]{revtex4-1}

\usepackage{graphicx}
\usepackage{dcolumn}
\usepackage{bm}
\usepackage{float}
\usepackage{color}

\def\n{{\bf n}}
\def\v{{\bf v}}
\def\J{{\bf J}}

\def\be{\begin{equation}}
\def\ee{\end{equation}}
\def\bea{\begin{eqnarray}}
\def\eea{\end{eqnarray}}

\begin{document}

\title{
Phase segregation of passive advective particles in an active medium
}

\author{Amit Das}
\affiliation{Simons Centre for the Study of Living Machines, National Centre for Biological Sciences (TIFR), Bangalore 560065, India.}

\author{Anirban Polley}
\affiliation{Raman Research Institute, C.V. Raman Avenue, Bangalore 560080, India.}

\author{Madan Rao}
\affiliation{Simons Centre for the Study of Living Machines, National Centre for Biological Sciences (TIFR), Bangalore 560065, India.}
\affiliation{Raman Research Institute, C.V. Raman Avenue, Bangalore 560080, India.}


\begin{abstract}
Localized contractile configurations or {\it asters} spontaneously appear and disappear as emergent structures in the collective stochastic dynamics of active polar actomyosin filaments.
 Passive particles which (un)bind to the active filaments get 
advected into the asters,
 forming transient clusters. 
We study the phase segregation of such passive advective scalars in a medium of 
dynamic asters, as a function of the
aster density 
and the ratio of the rates of aster remodeling to particle diffusion.
The dynamics of coarsening shows a violation of Porod behaviour, the growing domains have diffuse interfaces and low interfacial tension. The phase segregated steady state shows
strong macroscopic fluctuations characterized by multiscaling and intermittency, signifying rapid reorganization of macroscopic structures.
We expect these unique nonequilibrium features to manifest
in the actin-dependent molecular clustering at the cell surface.
\end{abstract}

\maketitle

Collections of  active  particles exhibit large concentration fluctuations, leading
 in many cases 
 to clumping and segregation into high and low density phases \cite{tonertu,marchettirmp} -
 the dynamics towards segregation and the space-time fluctuations of the order parameter, in single \cite{chate,Cates,Marenduzzo,Marchetti,baskaran1} and  two component  \cite{baskaran2,sriramsood} active systems, 
are found to be very different from their equilibrium counterparts.

 Recent studies on actomyosin-dependent molecular clustering at the cell surface \cite{Hancock,Goswami,Zanten,Lingwood,Kripacell} motivate us to explore a
  novel kind of phase segregation driven by activity.
These studies suggest that the 
cortical
layer of actomyosin generates active stochastic stresses
 and currents at the cell membrane, driving local clustering of several membrane components \cite{Goswami,Zanten,Kripacell}. 
 The spatiotemporal
 statistics of such clustering, driven by an active noise, is well described using an active hydrodynamics framework \cite{Kripacell}, where membrane components that (un)bind to actomyosin are 
 driven by the active filaments 
 - we will call these components passive advective scalars or simply {\it passive particles} \cite{Kripacell,AbhishekPNAS}. 
 The simultaneous presence of a 
 dense stationary actin meshwork at the cortex \cite{Morone} provides a 
 momentum sink  to the dynamics of the cell membrane components. Membrane curvature effects appear unimportant, 
 for such clustering occurs predominantly in the flat regions of the cell surface \cite{Goswami}.

Here we study the 
potential scale-dependent segregation of passive molecules from inert molecules (which do not bind to cortical actin) at the cell surface.
Our model system consists of passive and inert particles embedded in an active medium in two dimensions (2d). 
While we observe both {\it micro} and {\it macro} phase segregation upon varying the active noise strength
and
active filament concentration in the medium, 
the segregation dynamics
and the nature of the active phase segregated state 
are dramatically different from conventional phase segregation \cite{Bray}. 
First, the active phase segregation is driven by active advection and not by gradients of chemical potential, and so occurs at temperatures higher than the equilibrium segregation, a feature that is relevant to the cell surface context
 \cite{Baumgart,Ambika}. Second, 
the domain coarsening dynamics shows strong departures from Porod behaviour \cite{Bray}, which implies that the interfacial widths are diffuse 
with   {\it low interfacial tension}. 
Third, the active segregated phase shows strong fluctuations of the integrated order parameter and domain 
size, characterised by broad non-Gaussian distributions, in contrast
to the sharply peaked distribution in closed systems.
These strong fluctuations manifest as
{\it multiscaling} and  {\it intermittency} in the steady state, features seen 
in many driven nonequilibrium systems such as
turbulence \cite{Frisch}. As a result, the steady state exhibits a continual breakup and reformation of macroscopically large structures.
This {\it fluctuation dominated phase ordering} ({\it FDPO}) is similar to those observed in a variety of semiautonomously coupled dynamical systems, such as passive sliders on a fluctuating surface 
 \cite{passivesliders}, Burgers' fluids \cite{Burgers} and active nematics \cite{shraddha}.

{\it Dynamics of active fluids - asters as emergent structures} -- The basic hydrodynamic equations describing the polar contractile active fluid 
in terms of the active filament concentration $c$, the polar orientation $\n$, the apolar nematic orientation $\mathbf{Q}$ 
and the hydrodynamic velocity field $\v$ are described 
in recent reviews \cite{marchettirmp}. The precise form of the equations relevant to the present context are  
described in \cite{Kripacell,Kripaphase}.
Many studies, including \cite{Kripacell,Kripaphase} 
have shown that 
 the homogeneous, orientationally disordered configuration of active polar filaments is unstable to clumping and local orientation order to form spatially localized contractile regions -- these {\it emergent structures} \cite{baskaranemergent,kabir} have been referred to as {\em asters} \cite{Kripacell,Kripaphase}.

\begin{figure}[h]
\includegraphics[width=8.5cm,keepaspectratio]{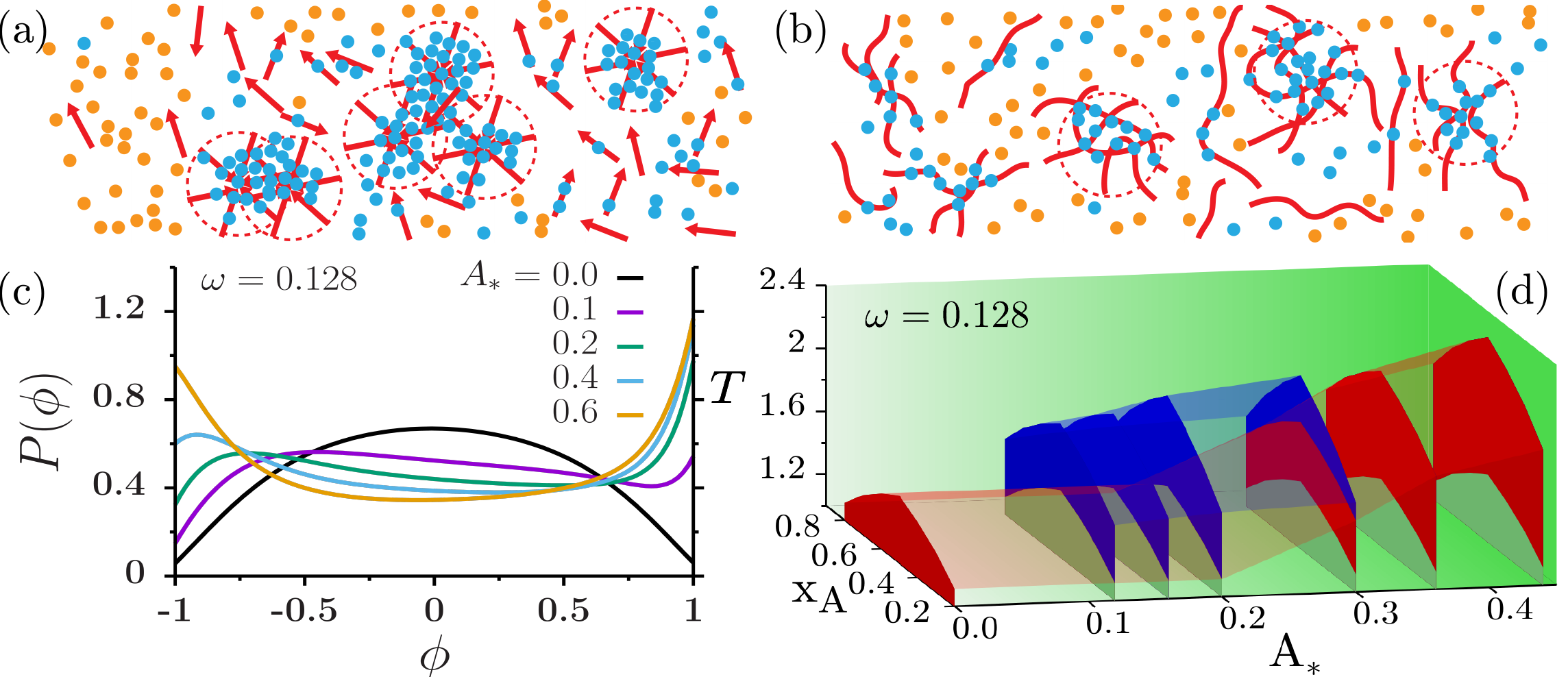}
\caption{\label{schematicnphase} 
(a-b) Schematic of a binary system of passive A (blue) and inert B (chrome) particles in an active medium. (a) The active medium consists of polar active filaments (red arrows) and {\it asters} (circled), within the asters the arrows point 
 inwards, while outside the asters they are orientationally uncorrelated. 
 (b) The active medium consists of
actin filaments (red) which form a mesh of actomyosin contractile platforms (circled), which draw in A particles that are bound to it.
(c) PDF of the order parameter $\phi$ for 
$\omega=0.128$ and different values of $A_{\ast}$ 
at a fixed temperature $T/T_c = 1.17$, where the equilibrium (Eqm) $P(\phi)$ shows a peak at $\phi=0$. The bimodal distribution for $\omega >0$ indicates phase segregation,
even at $T>T_c$. (d) Phase diagram in $(T, A_{\ast}, \mbox{x}_A)$ at $\omega=0.128$, shows micro (blue) and macro (red) phase segregation,
when $T>T_c$.}
\end{figure}

Introducing 
noise
parametrised by an ``active temperature'' $T_a$, results in 
a stochastic breakup and re-formation of asters (remodeling), with 
an exponential distribution of lifetimes \cite{Kripacell,Kripaphase,AbhishekPNAS} whose mean
$\tau_r$, is set by $T_a$.
This leads to a picture of a finite density of contractile structures or asters, whose
spatiotemporal fluctuations correspond to birth-death processes  \cite{AbhishekPNAS}.

Passive particles that (un)bind to the filaments are driven by these contractile flows; when bound they are 
advected with a current $\J \propto c \n$; when unbound, they move diffusively.
Thus the formation of contractile asters drive the localised clustering of passive particles (Fig.\ \ref{schematicnphase}a,b). The stochastic remodeling of asters give rise to dynamical fluctuations of clusters of passive particles 
explored in \cite{Kripacell}. 
Here, we 
 study the dynamics of phase segregation
of passive-inert particles moving in an active medium consisting of a finite density of remodeling asters - however,
rather than using the continuum hydrodynamic equations 
\cite{Kripacell,Kripaphase}, we work with 
an {\it effective theory}  that  incorporates all its relevant features.

{\it Effective dynamics using kinetic Monte Carlo} -- Consider a collection of 
contractile platforms, represented as discs of 
diameter $R_{\ast}$,
 distributed uniformly with area fraction $A_{\ast}$, in a 2d system of size $L$. 
In one model of the active fluid medium, 
 the filaments are distributed uniformly and point radially inward within the aster, 
 while they are oriented randomly outside the asters
 (Fig.\,\ref{schematicnphase}a). Alternatively, our model could describe a situation where the 
 active medium consists of 
 patches of contractile actomyosin mesh (Fig.\,\ref{schematicnphase}b). 
 In either case, 
the contractile platforms (e.g., asters) remodel with lifetimes chosen from an
exponential distribution $\exp[-t/\tau_r]$. 
In this dynamical background, consider a two component system of passive particles $A$ (at concentration, $\mbox{x}_A$) and inert particles $B$ (concentration, 
$\mbox{x}_B=1-\mbox{x}_A$).
The particles interact 
via a Lennard-Jones (LJ) short-range repulsion; in addition, $A-A$ and $B-B$ have a mutual attractive potential,
$u(r)=-J\, (\sigma/r)^{6}$,
 where $J$ is in units of temperature $k_BT$ and $\sigma$ is the LJ-length scale.

The passive
particles that bind/unbind to the active filaments with rates $\mathit{k_{b}}$ ($\mathit{k_{u}}$),  get advected 
with speed $v_0$ 
and timescale  $\tau_v = R_{\ast}/v_0$ in a direction pointing radially inwards, when the filaments are part of an aster. Outside the aster, the
passive particle moves diffusively, with a diffusion coefficient set by $v_0$ and the orientational decorrelation time of the filaments.

The kinetic Monte-Carlo (MC) dynamics\ \cite{suppinfo} involve (i) Kawasaki exchanges \cite{Binder,DavidLandau} of components $A$ and $B$ which obey 
detailed-balance, (ii) detailed-balance violating active advective updates whenever $A$ moves into an aster and (iii) detailed-balance violating stochastic birth-death
updates of asters with a lifetime distribution, $\exp[-t/\tau_r]$ \cite{note}. One MC step is $N$-attempts at exchange, $N$ being the total number of particles in the system. We have converted all time, length and energy units to s, $\mu$m and $k_BT$, respectively\ \cite{suppinfo}. We explore the dynamics
as a function of particle concentration $\mbox{x}_A$ and the active parameters characterizing the medium, namely $\omega = \tau_r^{-1}/\tau_d^{-1}$, the ratio of the aster remodeling rate to the
particle diffusion rate ($\tau_{d}$ is the diffusion time of particles over the scale of an aster), aster area fraction $A_{\ast}$ and the P\'eclet number 
$P_e \equiv \tau_v^{-1}/\tau_d^{-1}$.
We fix $P_e \approx187$ and duty ratio $K\equiv k_b/\left(k_b+k_u\right)=1$  \cite{suppinfo}. 

We find that the  steady state distribution is very sensitive to {\it both} active parameters,
$\omega$ and $A_{\ast}$.
At $\omega = 0$, the steady state is equivalent to particles diffusing in a quenched random media of stationary asters, characterized by an equilibrium distribution.
At the other extreme,
$\omega \gg 1$, the asters remodel much faster than the particle movement,
 the steady state is essentially that of equilibrium diffusing particles. It is only for
 $0 < \omega \ll 1$, when the particles move fast compared to aster remodeling, that we get a genuine
 nonequilibrium steady state, where clusters form, fragment and
move towards newly formed asters.

\begin{figure}[h]
\includegraphics[width=8.5cm,keepaspectratio]{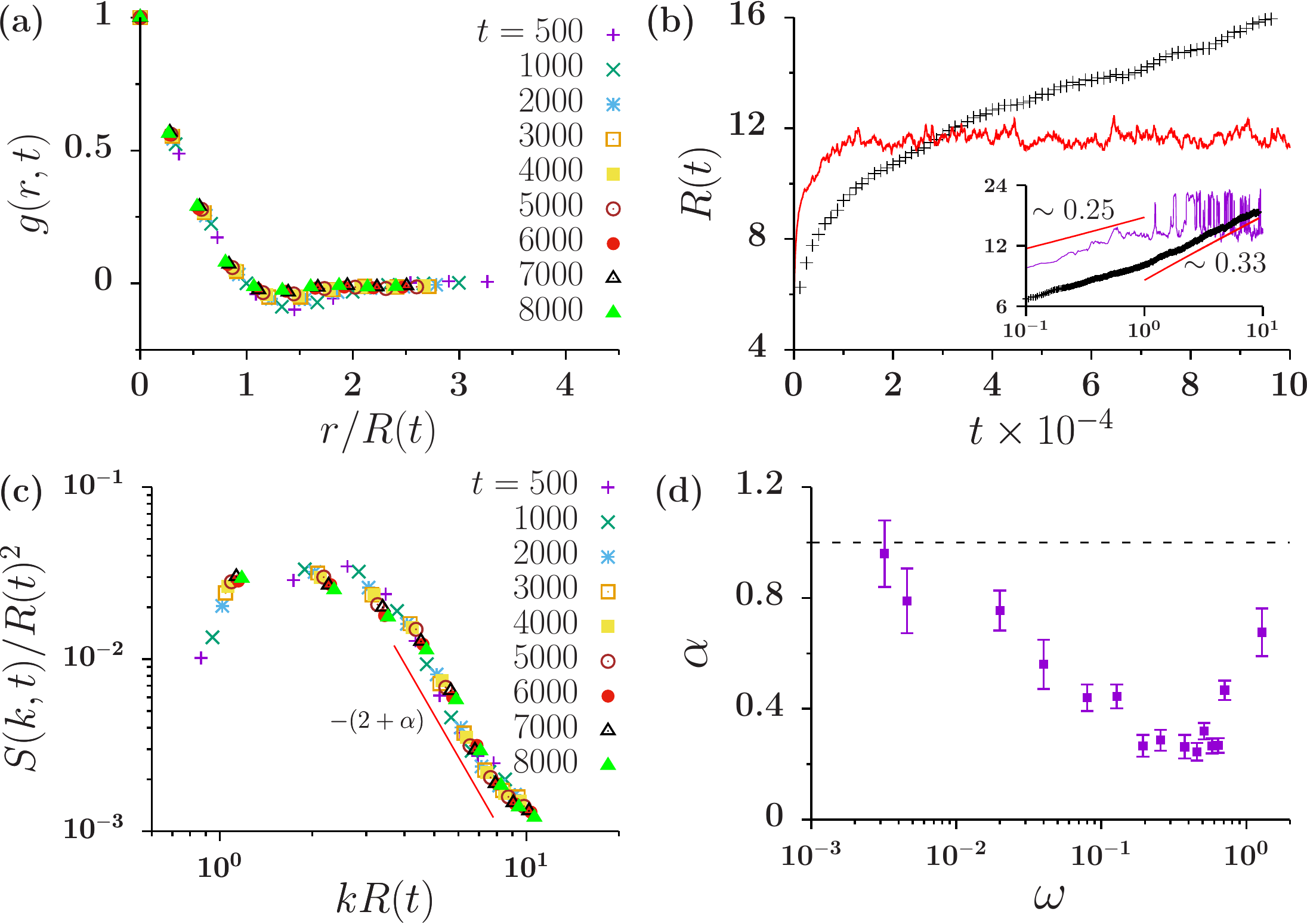}
\caption{\label{corr-func}
Active coarsening dynamics, shown here for $\omega = 0.128, A_{\ast}=0.3$ and $T/T_c = 1.17$ : (a) Dynamical scaling of the order parameter correlation function $g(r,t)$ with domain size  $R(t)$.  The finite intercept of $g(r,t)$ at the scaled $r/R(t) \to 0$ is an indication of
long-range order.
(b) $R(t)$ grows with time (red) and saturates to a value that depends on $A_{\ast}$ and $\omega$ \cite{suppinfo}, unlike the 
 equilibrium phase segregation at $T<T_c$ (`+' symbols), where the domain size keeps growing as $t^{1/3}$ (black symbols in inset) until it reaches system size.
In contrast, for active phase segregation, 
$R(t) \sim t^{1/4}$ before saturating (purple line), here $\omega = 0.128, A_{\ast}=0.6$.
 (c) Structure factor $S(k,t)$ also exhibits dynamical scaling but a violation of Porod behaviour - $S(kR(t)) \sim k^{-(2+\alpha)}$ at large $k$, where
$0< \alpha \leq 1$.
(d)  $\alpha$ varies non-monotonically with $\omega$, showing ``equilibrium'' behaviour (represented by
the dashed line) at $\omega=0$ and $\omega \gg 1$. }
\end{figure}

{\it Phase diagram} -- We first obtain the equilibrium phase diagram in the $\left(T-\mbox{x}_A\right)$ plane and the critical temperature $k_B T_c / J = 1.43$, by monitoring the
probability distribution function (PDF) $P(\phi)$ of the coarse-grained segregation order parameter,
$\phi=\frac{\rho_A-\rho_B}{\rho_A+\rho_B}$, where $\rho_A, \rho_B$ are the local densities of $A$ and $B$, respectively. The order parameter $\phi$ is coarse-grained over a scale $3\sigma$ which is smaller than the typical domain size \cite{suppinfo,Binder}. The phase boundary is determined by the appearance of a distinct double peak in $P(\phi)$ at
$\phi= \pm 1$.

 We next
determine the nonequilbrium phase diagram, as a function of $\left(T, A_{\ast},\mbox{x}_A \right)$
at different values of $\omega$.
Unlike the equilibrium case, phase segregation here is driven by the active advection 
of one of the components into asters, rather than by gradients in chemical potential, which
 implies that segregation can occur even at $T > T_c$. Its extent depends on $A_{\ast}$ - for small $A_{\ast}$ only micro-phase segregation takes place
(the domain sizes are roughly the aster size); as $A_{\ast}$ increases to a finite fraction of the system size,
we start getting macro-phase segregation. Figure\,\ref{schematicnphase}c,d indicates that the nature of segregation
 depends on the values of $\omega$ and $A_{\ast}$. 

{\it Coarsening dynamics} -- We  next study the dynamics of segregation following a quench from the high temperature mixed phase to a
region between the active and equilibrium phase boundaries (red and blue shaded regions in Fig.\,\ref{schematicnphase}d) 
- a situation relevant 
to the cell membrane, 
where equilibrium lipid phase segregation has only been observed at temperatures much lower than physiological temperatures
 \cite{Baumgart,Ambika}. 
We study the dynamics of the order
parameter correlation function $g(r,t) = \langle \phi(0,t) \phi(r,t)\rangle$ as the system coarsens. As in conventional phase ordering  \cite{Bray}, $g(r,t)$ appears to obey
dynamical scaling (Fig.\,\ref{corr-func}a, Fig.\,S2), $g(r,t)=g(r/R(t))$, where the domain size $R(t)$ is obtained from the first-zero of $g(r,t)$.
The mean domain size 
$R(t)$ grows before saturating to a value that depends on $\omega$ and $A_{\ast}$ -  a power law fit yields (Fig.\,\ref{corr-func}b, Fig.\,S3), $R(t) \sim t^{1/4}$, 
different from the conventional phase ordering growth of $t^{1/3}$ \cite{Bray}. Note that the maximal domain size is much smaller than the equilibrium case, even though it increases
with increasing system size $L$ (with $A_{\ast}$ held fixed). The fourier transform of $g(r,t)$, the structure factor $S(k,t)$ also exhibits scaling $S(k,t)=[R(t)]^2\,S(kR(t))$ (Fig.\,\ref{corr-func}c, Fig.\,S4). In conventional phase segregation,
$S(kR(t))$ at large $k$ goes as $k^{-3}$, known as Porod's law \cite{Bray}, signifying that the interface separating the growing phases is sharp. Here, in contrast, the tail of $S(kR(t))$ appears to go as $k^{-(2+\alpha)}$, 
where $\alpha$ depends on $\omega$ and
lies in the range $0 < \alpha \leq 1$ (Fig.\,\ref{corr-func}d). This departure from Porod signifies that the interfaces 
are diffuse with very low interfacial tension, the interfacial width being $\ell \sim R^{1-\alpha}$ (while still having a well-defined ``domain'', $\ell/R \ll 1$ for large $R$).
This also implies that $g(r/R(t))$ has a {\it cusp} nonanalyticity at small $r/R(t)$, going as $g(y) \sim 1-y^{\alpha(\omega)}+ \ldots$.

\begin{figure}
\includegraphics[width=8.5 cm,keepaspectratio]{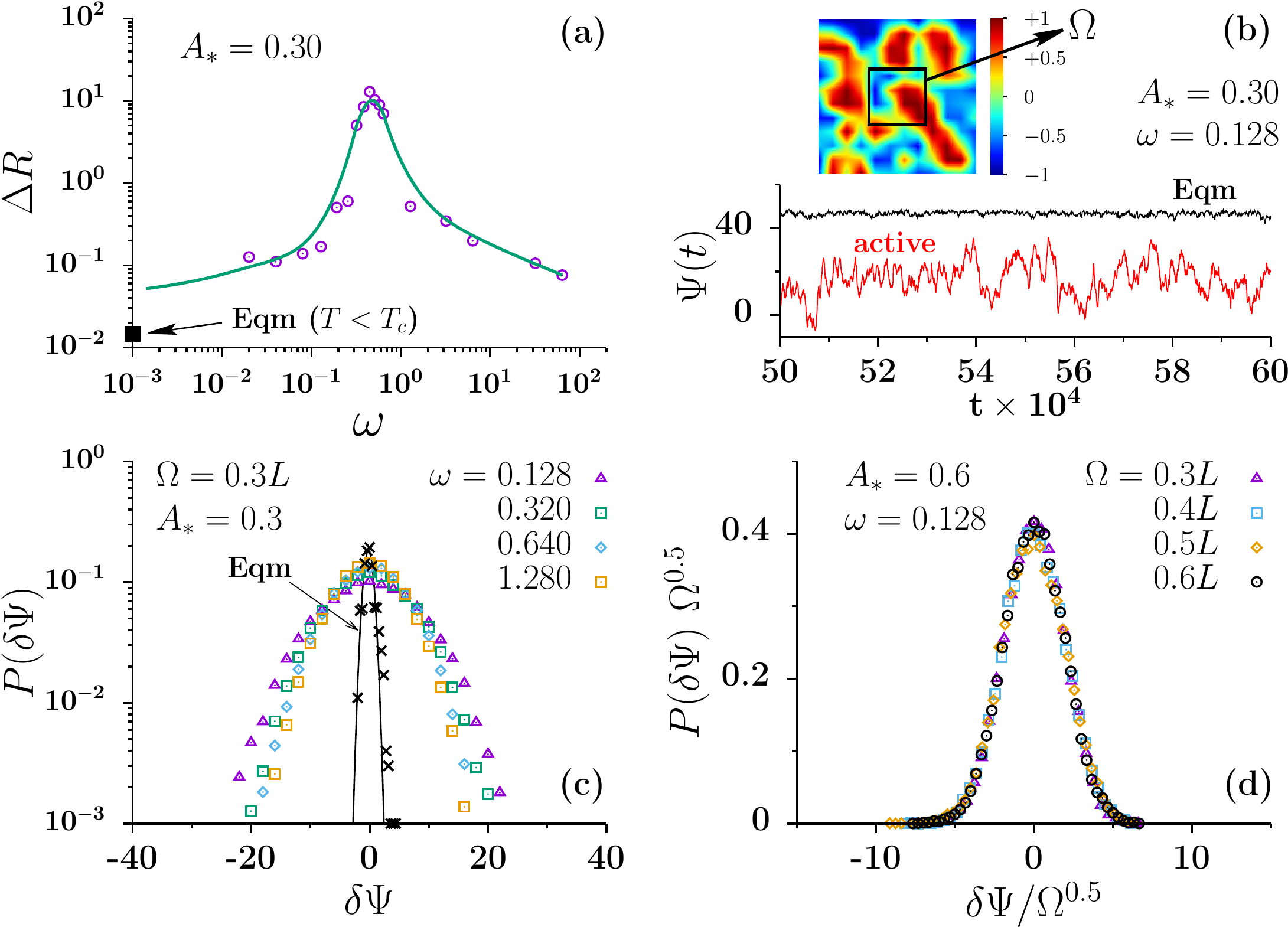}
\caption{\label{time-series}
Time series of statistical quantities show large fluctuations in the phase segregated steady state.
(a)  Variance of $R(t)$, $\Delta R(t) \equiv \langle(R-\langle R\rangle)^2\rangle$, shows a non-monotonic dependence on $\omega$ (open circles). Line through data points is a piece-wise cubic spline fit. For comparison we show $\Delta R(t)$ for the equilibrium segregation state when $T<T_c$ (filled square). (b) Time series of region-wise order parameter $\Psi$, evaluated over the square patch of size $\Omega$ (inset), for equilibrium and active segregation states. Color-bar indicates the value of the local order parameter $\phi$. The mean value of  $\Psi$ is smaller and the fluctuations are larger in the active state.
(c) PDF  $P(\delta \Psi)$, where 
$\delta \Psi = \Psi - \langle \Psi \rangle$, at different $\omega$'s for fixed $A_{\ast}$, shows a much broader width than that for equilibrium segregated state. (d)  $P(\delta \Psi)$ shows scaling with the region-size $\Omega$,
the collapse curve is a gaussian. Data displayed for $T/T_c = 1.17$.
}
\end{figure}

{\it Fluctuation dominated phase ordering} -- The most dramatic aspect of the active segregated steady state is in the nature of fluctuations of a variety of statistical quantities.
{\it Supplemental Movies} \cite{suppinfo} show that the dynamics of $\phi$-profile is characterized by strong fluctuations both at the interface and bulk, unlike the equilibrium situation where the fluctuations are appreciable 
only at the interface. To get a quantitative handle, we study the time series of $R$, the 
region-wise order parameter, $\Psi = \int_{\Omega} \phi ({\bf r})\, d{\bf r}$ (where the integration is done over a large enough region of size $\Omega \ll L$), and the fourier amplitude of $\phi$ at low $k$ (Fig.\,S5-S7), $\Phi(1)\equiv\vert\phi(k=2\pi/L)\vert$, $\Phi(2)\equiv\vert\phi(k=4\pi/L)\vert$, $\Phi(3)\equiv\vert\phi(k=6\pi/L)\vert$, $\ldots$.
We find that
the corresponding PDFs at steady state, show 
large variance and heavy non-Gaussian tails (Fig.\,\ref{time-series}, Fig.\,S5).
Figure\,\ref{time-series}d shows that the PDF $P(\delta \Psi) \sim \frac{1}{\sqrt{\Omega}} \exp\left[-\Omega\,A(\omega) \left(\delta \Psi/\Omega\right)^2\right]$, expressed in the large deviation function form \cite{largedeviation},
where $\delta \Psi=\Psi - \langle \Psi\rangle$. 
These observations are consistent with other driven non-equilibrium systems, such as passive sliders on
fluctuating interfaces \cite{passivesliders} and Burgers' fluids \cite{Burgers}. 

\begin{figure}
\includegraphics[width=8.5 cm,keepaspectratio]{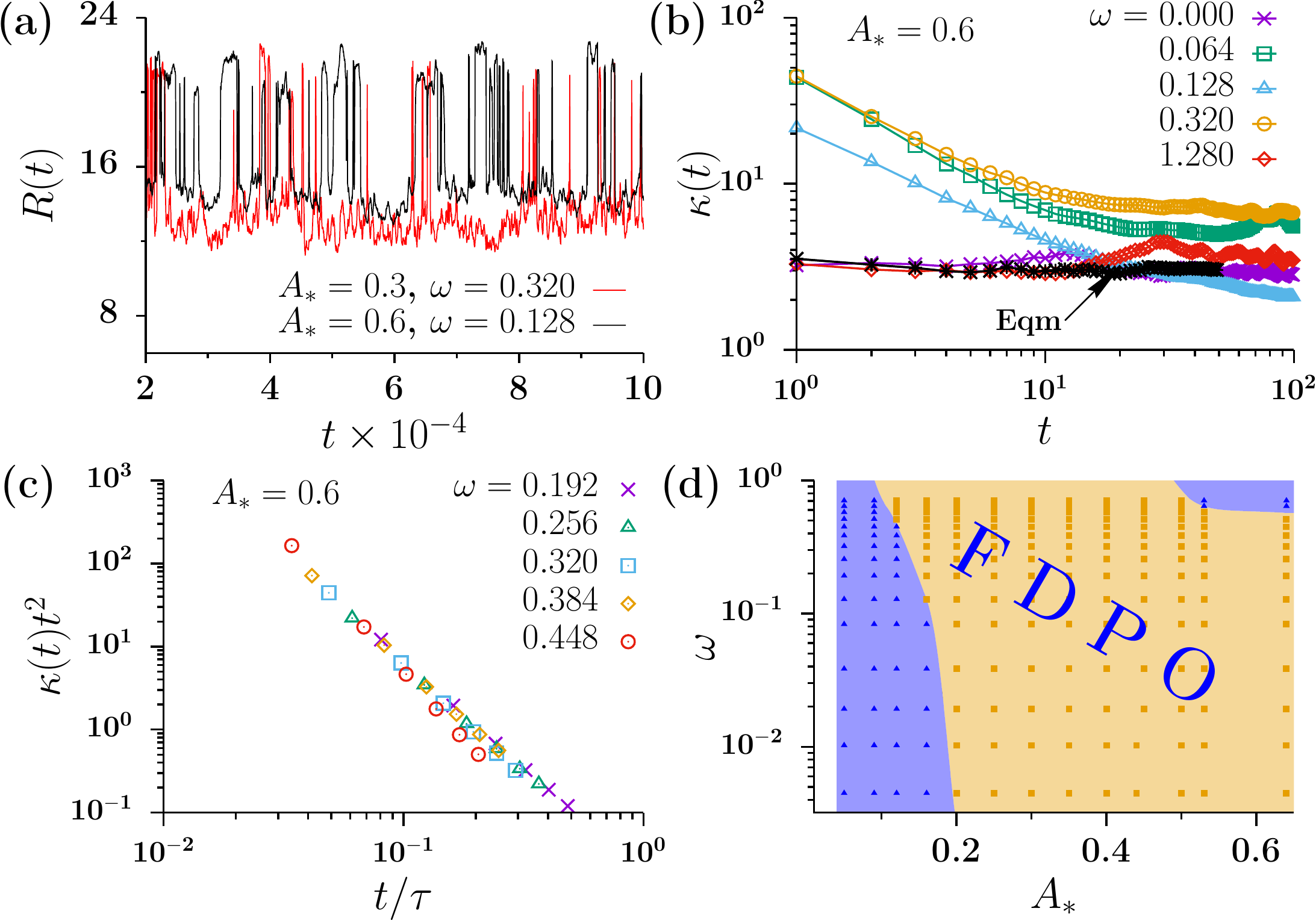}
\caption{\label{multiscalingnfdpophase} 
Strong fluctuations in steady state characterized by intermittency and multiscaling.
(a) Time series of domain size $R$ for different values of $\omega$ and $A_{\ast}$ shows sudden precipitous drops and rises, suggesting {\it
intermittent} behaviour.
(b) Kurtosis $\kappa(t)$ as a function of $t$, for different values of $\omega$. (c) Scaling collapse of
$\kappa(t)$ with scaling variable $t/\tau$, where $\tau = \tau_r \left(\omega^2 + \omega_0^2\right)$ and the numerical parameter $\omega_0=0.1$
\cite{camacho} at $A_{\ast}=0.6$, shows divergence as $t/\tau \to 0$.
(d) Region (chrome) in $\omega-A_{\ast}$ plane which exhibits fluctuation dominated phase ordering (FDPO). There is no macroscopic phase ordering in the blue regions.
 Data displayed for  $T/T_c = 1.17$.
}
\end{figure}


The time series of various statistical quantities $X(t)$ alternates between periods of quiescence
and large changes over very short time scales (Fig.\,\ref{multiscalingnfdpophase}a), indicating a  breakdown of self-similarity due to intermittency. 
We characterize this behaviour in terms of the $n$-point {\it structure functions} $S_n = \langle \left(X(t) - X(0)\right)^n \rangle$, where
$\langle \ldots \rangle$ is an average over histories, as in studies of turbulent systems \cite{Frisch}. Here we compute such structure functions for 
$R(t)$.

These structure functions are expected to scale with time as $S_n \sim t^{n \gamma_n}$ (Fig.\,S8).
 In equilibrium phase
segregation (away from critical points), the statistical quantities are self-similar, implying $\gamma_n=\mbox{const}$
 - called {\it gap scaling} (Fig.\,S8).
 In the active phase segregation,
$\gamma_n$ is a nonlinear function of $n$ - a phenomenon known as {\it multiscaling} (Fig.\,S8).
This breakdown of self-similarity is a 
consequence of intermittency (which is dominated by extreme events), as seen by the behaviour of 
``flatness'' or kurtosis, $\kappa(t) = S_4(t)/S_2^2(t)$ (Fig.\,\ref{multiscalingnfdpophase}b). 
Intermittence shows up as a divergence in $\kappa$ 
as $t/\tau \to 0$ (Fig.\,\ref{multiscalingnfdpophase}b), where $\tau$ is a time scale
which characterizes the lifetime of the largest structures in
the system \cite{Himani}. We argue that $\tau$ scales as the remodeling time $\tau_r$ for 
large $\tau_r$ 
and is set by $\tau_d$
for 
small $\tau_r$ -
 this suggests the following nonlinear form, $\tau=\tau_r \left[ \omega^2 + \omega_0^2\right]$, where $\omega_0$ is a
numerical fit parameter \cite{camacho} (Fig.\,\ref{multiscalingnfdpophase}c).
The onset of divergence appears earlier for faster remodeling $\tau_r$. Note that $\kappa(t)$ shows ``equilibrium-like'' non-intermittent behaviour when $\omega = 0$ and for $\omega > 1$ (Fig.\,\ref{multiscalingnfdpophase}c). This is in agreement with our earlier assertion that these limiting cases correspond to steady states of an
equilibrium model.
%

Non-Porod behaviour, non-Gaussian distributions, multiscaling and intermittency are signature features of  
 a driven nonequilibrium steady state called the {\it fluctuation-dominated phase ordered (FDPO) state} \cite{passivesliders}. The most 
 dramatic manifestation of the FDPO, however, appears in the continual remodeling of the largest scale structures at steady state ({\it Movie S2}). 
 This visual feature is captured by the time series of the dominant fourier amplitudes of the order parameter $\Phi(m)$, for $m=1,2,3, \ldots$. The breakup of the
 larger structure into 2 or 3 domains is reflected in the {\it anticorrelation} of the cross-correlation function $\langle \Phi(1,0)\Phi(2,t)\rangle$ and $\langle \Phi(1,0)\Phi(3,t)\rangle$ (Fig.\,S6, S7) - 
 the  decrease in the amplitude $\Phi(1)$ feeds into the amplitudes $\Phi(2)$ and $\Phi(3)$ \cite{kapribarma}. In Fig.\,\ref{multiscalingnfdpophase}d, we depict the region
 in the steady state phase diagram in $\omega-A_{\ast}$, where FDPO obtains.


{\it Discussion} -- In this paper we have studied the dynamics of phase segregation of passive and inert particles in the background of a fluctuating active medium in 2d. 
We find that the active driving can induce segregation 
at $T>T_c$, the equilibrium transition temperature.
The dynamics of coarsening, described by domain growth and sharpness of growing interfaces is very distinct from its equilibrium counterpart. More
dramatic is the nature of the phase segregated steady state, which shows macroscopic fluctuations of the order parameter, multiscaling and intermittency, characteristics of
a {\it FDPO} state (Fig.\,\ref{multiscalingnfdpophase}d).


We have now extended this study to temperatures $T<T_c$, where the system can undergo an equilibrium phase segregation in the absence of active driving. Here again, we find that turning on activity results in a breakdown of conventional segregation dynamics - $R(t)$ grows and saturates to a value determined by $A_{\ast}$ and $\omega$ and the segregated 
phase shows FDPO. Activity thus destroys the very large domains obtained in equilibrium phase segregation and makes them more dynamic and intermittent.
While our motivation and focus here has been on the segregation of passive particles driven by actomyosin, the 
general features of our study should be applicable to other hybrid passive/active systems such as particles driven by 
motor-microtubule complexes \cite{microtubules}.
We look forward to testing many of the predictions reported here  in the scale-dependent actomyosin-based segregation of molecules at the cell surface \cite{Suvrajit}
and  in {\it in-vitro} realizations of supported membranes attached to a thin active layer of actomyosin \cite{DKoster}.

We thank S. Saha, D. Koster and S. Mayor, and especially K. Vijaykumar, A. Dhar, M. Barma and D. Das for insightful discussions.


\begin{thebibliography}{30}
\bibitem{note} Since the birth-death statistics of asters parametrized by $\tau_r$ is controlled by an active temperature $T_a$, this makes it akin to
a two-temperature model which is inherently nonequilibrium.
\end{thebibliography}
\end{document}